# On Photon Spin and the Electrodynamic Origin of the Charge of the Electron

Ulrich C. Fischer

*Interface Physics Group; Physikalisches Institut, Westfaelische Wilhelms-Universitaet, Muenster, Germany*

*e-mail: fischeu@wwu.de*

**Abstract**

We recently performed experiments on the transfer of photon spin to electron orbital angular momentum. For an interpretation of the experimental results we used a classical electrodynamic model of the photon as a propagating electromagnetic solitary wave which is developed in detail here. A linearly polarized monochromatic photon is considered as a propagating solitary electromagnetic wave of finite energy $h\nu$ which carries an angular momentum $\frac{h}{2\pi}$ with the frequency $\nu$ and Planck's constant h. This model has, apart from being a tool for an interpretation of our experimental results, far reaching consequences of fundamental relevance and guides us to an outline to a unified quantum theory of electromagnetism and gravitation including an explanation of the electrodynamic origin of the quantized charge of an electron.

**Introduction**

We recently performed experiments on the transfer of photon spin to electron orbital angular momentum [1]. For an interpretation of the experimental results we developed a classical electrodynamic model of a photon as a propagating electromagnetic solitary wave which is elaborated here .

Photons can be regarded as the electromagnetic radiation which is emitted from an oscillating electric dipole. Let us consider monochromatic photons of a defined energy $U = h\nu$ (h is Planck's constant and $\nu$ is the frequency of the photon) which are emitted from a z-oriented oscillating electric dipole. From the point of view of quantum theory, photons are bosons which possess a spin 1 oriented in the direction of the emitting dipole and an associated angular momentum $\frac{h}{2\pi}$ [2], where h is Planck's constant. The dipole radiation is in the transverse x-y directions and not in the z-direction and the radiation in the transversal radial direction is linearly polarized with the polarization in the z direction. From the point of view of electrodynamics, linearly polarized radiation carries no spin. So there is a discrepancy between the quantum theoretical picture, where the linearly polarized emission of electromagnetic energy h$\nu$ carries a spin 1 and from a point of view of electromagnetic theory, where the same dipole radiation carries a spin 0. This discrepancy was reconciled in a thought experiment by Raman and Bhagavantam, who experimentally proved that linearly polarized photons have a quantized angular momentum [2]. Consider a beam of linearly polarized light which may be considered as a current of photons. If you do an experiment which is able to detect the spin you would find an average of spin 0 in this experiment. If you now attenuate the beam such that single photons can be discriminated with a spin sensitive detector you would then detect photons with an equal probability of spin -1 and spin +1 with an average of spin 0. So they argue that the spin of a linearly polarized photon is a quantum mechanical property which does not exist for a classical linearly polarized beam of electromagnetic radiation. They referred in their argumentation to the book "Principles of Quantum Mechanics" of P. A. M. Dirac [3]. But referring to this book, it seems, that Dirac had a different view. He defines the photons emitted from an electric dipole transition as linearly polarized photons. It is possible to detect photons emitted from a dipole in a longitudinal direction e.g. by collecting the longitudinal radiation with a lens as shown schematically in fig. 1.

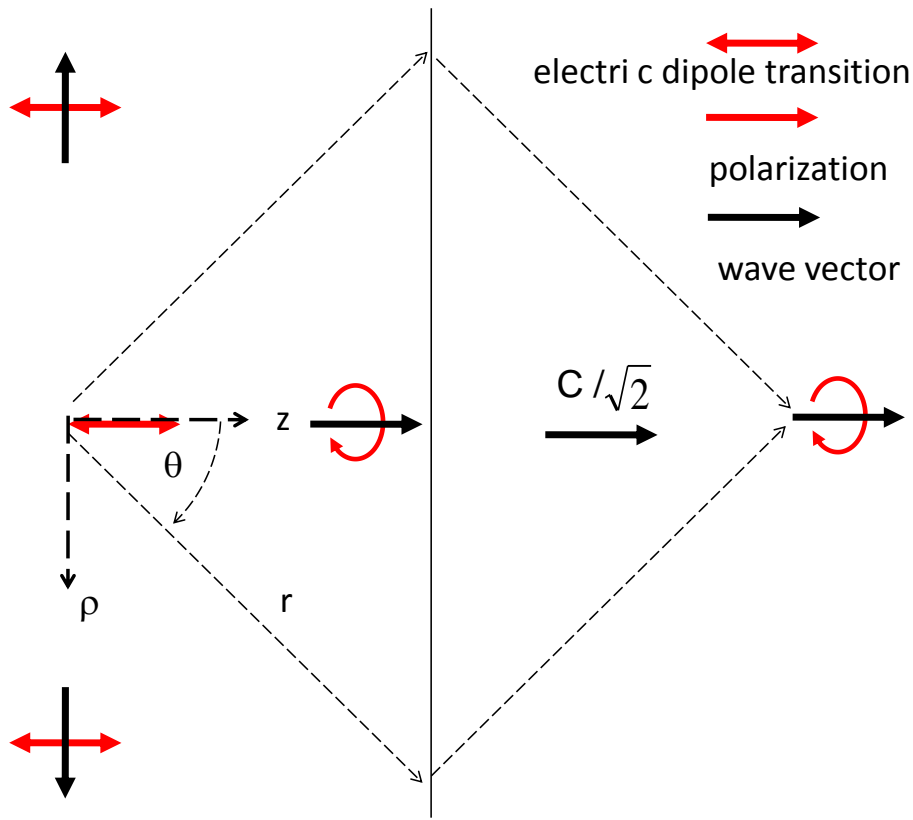

**Fig. 1** *Schematic view of a photon emitted by an electric dipole transition propagating in the longitudinal direction, which is focused by means of a lens. The center of field energy of the photon propagates in the longitudinal direction at a speed of $\frac{c}{\sqrt{2}}$. The radiation in the longitudinal direction is right handed circularly polarized, whereas the emission in the transverse directions is linearly polarized.*

According to Dirac, the light emitted from a dipole in the longitudinal direction has a right handed circular (rc) polarization. It is well known from electromagnetic theory, that a rc polarized beam of light carries a spin in the direction of propagation. So the beam of light of an energy $h\nu$ emitted from a dipole in the longitudinal direction, which can be regarded as a linearly polarized photon, carries a spin 1 and an angular momentum $h/2\pi$. According to Dirac, a time reversed propagating beam with negative frequency has a lc polarization and carries a spin angular momentum $-\frac{h}{2\pi}$ in the longitudinal direction of propagation. A solution to the discrepancy comes from the fact that a linearly polarized stationary beam of light carries no spin in accord with the well known property of linearly polarized light which can be considered as a superposition of a right and left circular polarized light with a phase difference of $\pi$ and thus the spin of the stationary polarized beam is 0. So the argument of Raman and Bhagavantam is correct for a stationary linearly polarized beam of light and indeed, in the stationary field distribution of an emitting electric dipole, which corresponds to a stationary radially polarized beam, which extends from -∞ to +∞ in time we find no signature of a spin. But a monochromatic stationary electromagnetic field of an oscillating dipole cannot be considered as the field arising from a single photon, because a stationary field requires a constant input of energy and its energy content is infinite. The situation changes for a propagating beam of light with a positive frequency. We show here with simple arguments, that the electromagnetic field of a linearly polarized beam of light of the energy hν of a single photon emitted from a light emitting dipole away from its origin carries in its wave front an angular momentum h/2π in the z- direction of the dipole and the longitudinal direction of propagation by looking at the spatial distribution of the electric and magnetic components of its radiation field.

But first we extend the concept of the photon of positive frequency, which, by Dirac, was considered as the quantized emission of an electromagnetic energy of spin 1 due to an electric dipole transition. Photons can, however, also be generated by a magnetic dipole transitions since oscillating magnetic dipoles generate polarized radiation fields which are indistinguishable from the radiation fields generated by an electric dipole oscillation as long as one has no information about the source of radiation. We define a photon emitted by an electric dipole transition as an electric photon and a photon emitted by a magnetic dipole transition as a magnetic photon respectively. It turns out that the emission of electric photons have a right handed circular polarization in the longitudinal propagation direction whereas magnetic photons have a left handed circular polarization, respectively. We limit our considerations to electric and magnetic photons. The concept can be extended to photons with multipolar radiation fields according to Davydov [4]

We first consider the far field distribution of the electric field E and the magnetic field H of monochromatic electric dipole radiation according to Stratton [5] for a dipole oriented in the z direction of a spherical coordinate system as illustrated in fig. 1 :

$$E_\theta(r,t) = -\frac{k^2}{4\pi\varepsilon_0 r}\sin\theta |p|\, e^{-i\omega t^*}$$

$$H_\varphi(r,t) = -\frac{k\omega}{4\pi r}\sin\theta |p|\, e^{-i\omega t^*}$$

The fields for a propagating photon are obtained by putting for t>0 $t^* = t - \frac{r}{c} = 0 \Rightarrow r = ct$.

For a forward propagating photon in the longitudinal z direction with $z > 0$ and $r = \sqrt{z^2 + \rho^2}$ we obtain

$$(1) \quad E_\theta(r) = -\frac{k^2}{4\pi\varepsilon_0 r}\sin\theta |p|$$

$$(2) \quad H_\varphi(r) = -\frac{k\omega}{4\pi r}\sin\theta |p|$$

Similar expressions hold for the field distribution of a magnetic dipole:

$$H_\theta(r,t) = -\frac{k^2}{4\pi r} \sin\theta |m| e^{-i\omega t^*}$$

$$E_\varphi(r,t) = \frac{k^2}{4\pi r} \sqrt{\frac{\mu_0}{\varepsilon_0}} \sin\theta |m| e^{-i\omega t^*}$$

and again for t>0, $z > 0$ putting $t^* = t - \frac{r}{c} = 0 \Rightarrow r = ct$,

$$(3) \quad H_\theta(r) = -\frac{k^2}{4\pi r} \sin\theta |m|$$

$$(4) \quad E_\varphi(r) = \frac{k^2}{4\pi r} \sqrt{\frac{\mu_0}{\varepsilon_0}} \sin\theta |m|$$

**Electromagnetic properties of a magnetic photon**

Let us first consider the magnetic photon in the reference frame propagating at a speed of $\frac{c}{\sqrt{2}}$ in the longitudinal positive z-direction. The longitudinally propagating photon can be considered as a longitudinally stationary and laterally expanding electromagnetic object. From the field distributions (3),(4) we can derive a magnetic surface charge and an angular momentum of the propagating wave front. The magnetic field $H$ has the dimension of a magnetic polarization density or magnetic surface charge density $\sigma_m$. We can derive a magnetic surface charge $q_{m1}$ by integrating the magnetic field over the hemispherical wave front of the photon propagating in the positive longitudinal z- direction by using Gauss's law [5] for the magnetic field $\vec{H}$ :

$$\oint \sigma_m \, da = \int \nabla \cdot \vec{H} \, dV = q_{m1}$$

Using spherical coordinates we obtain with

$$\nabla \cdot \vec{H} = \frac{1}{r\sin\theta} \frac{\partial}{\partial\theta} \sin\theta H_\theta$$

and inserting $H_\theta$ from (3) :

$$\nabla \cdot \vec{H} = -\frac{k^2 |m|}{4\pi r^2 \sin\theta} \frac{\partial}{\partial\theta} \sin^2\theta = -\frac{k^2 |m| \cos\theta}{2\pi r^2}$$

$$(5) \quad q_{m1} = \int \nabla \cdot \vec{H} \, dV = -|m| \int \frac{k^2 \cos\theta}{2\pi r^2} r^2 d\theta \, d\varphi \, \delta(r-r')dr' = -k^2 |m|$$

In this expression $\delta(r'-r)$ is Dirac's delta function. A further magnetic surface charge density can be derived from an electric displacement current, which is associated to the azimuth al electric field $E_\varphi$ of the magnetic photon. Using relation (4) an azimuthal displacement current density $j_{l\varphi}$ can be defined:

$$j_{l\varphi} = \varepsilon_0 \frac{\partial E_\varphi}{\partial r} \frac{dr}{dt} = \varepsilon_0 c \frac{\partial E_\varphi}{\partial r} = -\frac{k^2}{4\pi r^2} sin\theta |m|$$

A total azimuthal current can be derived:

$$(6) \quad J_\varphi = \iiint_{\theta=0, \varphi=0, r'=0}^{\frac{\pi}{2}, 2\pi, \infty} \frac{k^2}{4\pi r^2} sin\theta |m| r^2 \delta(r-r') d\varphi \, d\theta \, dr' = \frac{k^2}{2} |m| = q_{m2}$$

This circular displacement current can be considered as a magnetic surface charge of opposite polarity. For the total magnetic surface charge we obtain:

$$(7) \quad q_m = q_{m1} + q_{m2} = -\frac{k^2}{2}|m|$$

We conclude, that the magnetic photon carries a magnetic surface charge which corresponds to a magnetic monopole. The center of magnetic charge of the monopole propagates at a speed of $\frac{c}{\sqrt{2}}$, i.e. the speed of a surface wave.

There is also an angular momentum associated with the rotary displacement current. We first derive a radial linear momentum density, which is associated with the Poynting vector:

$$p_r = \frac{1}{c^2}|E_\varphi \times H_\theta| = -\frac{k^4}{16\pi^2 r^2 c^2} \sqrt{\frac{\mu_0}{\varepsilon_0}} sin^2\theta |m^2|$$

for the local time derivative of this momentum we obtain:

$$\frac{d}{dt} p_r = \frac{d}{dr} p_r \cdot \frac{dr}{dt} = c \frac{d}{dr} p_r = \frac{k^4}{8\pi^2 r^3 c} \sqrt{\frac{\mu_0}{\varepsilon_0}} sin^2\theta |m^2|$$

$$\rightarrow \frac{d}{dt} p_r = -\frac{2c}{r} p_r$$

This local time derivative of the linear momentum can be considered as a radial inertial force, which is related to the longitudinal propagation of the photon. This radial inertial force has a cylindrical ρ-component perpendicular to the z-axis

$$F_\rho = -sin\theta \frac{d}{dt} p_r \vec{n}_\rho$$

This cylindrical component is a centripetal force which leads to the azimuthal displacement current. In this context a local azimuthal torque $M_\varphi$ can be defined:

$$\vec{M}_\varphi = \vec{r} \times \vec{F}_\rho = -r sin^2\theta \cdot F_\rho \vec{n}_\varphi$$

This torque can be related to a time derivative of a local angular momentum:

$$\vec{M}_\varphi = \frac{d\vec{L}_{loc}}{dt} = \vec{\omega} \times \vec{L}_{loc}$$

By volume integration we obtain an expression for the total angular momentum L

$$\omega L = \iiint_{\theta=0,\varphi=0,r'=0}^{\frac{\pi}{2},2\pi,} r\sin^2\theta \frac{d}{dt} p_r r^2 \, \delta(r'-r) d\varphi d\theta dr'$$

$$= \frac{k^4}{8\pi^2 c}\sqrt{\frac{\mu_0}{\varepsilon_0}}|m^2| \int_{\varphi=0}^{2\pi}\int_{\theta=0}^{\frac{\pi}{2}} \sin^4\theta \, d\varphi d\theta$$

$$\omega L = \frac{k^4 \mu_0}{32}|m^2|$$

(8) $\quad L = \frac{k^4 \mu_0}{32\omega}|m^2|$

Considering that the angular momentum of the photon is quantized, we obtain from relation (7), by setting $L = \frac{h}{2\pi}$, the result:

(9) $\quad |m^2| = \frac{32\, h\nu}{\mu_0\, k^4}$

And thus we obtain

(10) $\quad q_m^2 = \frac{2\, h\nu}{\mu_0}$

We have thus shown, that the propagating magnetic photon can be considered as a solitary wave with a magnetic pole of a magnetic surface charge $q_m = \sqrt{\frac{2h\nu}{\mu_0}}$ and an angular momentum $\frac{h}{2\pi}$. The magnetic surface charge is equivalent to a magnetic monopole. The center of the magnetic surface charge of the soliton propagates at a speed $\frac{c}{\sqrt{2}}$ in the longitudinal direction. In the inertial frame of its longitudinal propagation a magnetic photon can be considered to have a rest mass $m_{0;photon}$ and we obtain the relations :

(11) $\quad m_{0;photon} = \frac{h\nu}{c^2};\qquad$ (16) $\quad \lambda = \frac{h}{m_{0,photon}\, c}$

We remark, that the rest mass of the photon is related to its wavelength in the same way as the rest mass of the electron $m_{0;electron}$ is related to the quantum mechanical particle wavelength of the electron, the Compton wavelength $\lambda_c$:

$$\lambda_c = \frac{h}{m_{0;elctron}\, c}$$

The quantized magnetic monopole of the magnetic photon is associated with a magnetic potential energy which has the same physical dimension as a torque [Nm]

$$(12) \quad M = \frac{1}{2}\mu_0 q_m^2 = h\nu$$

**Electromagnetic properties of an electric photon**

In close analogy to the derivation of the magnetic surface charge for the magnetic photon we can derive an electric surface charge or an equivalent dipole moment of the electric photon

$$(13) \quad q_e = -k^2|p|$$

Also for the electric photon we can derive an angular momentum. For the density of the linear momentum we obtain:

$$p_r = \frac{1}{c^2}|E_\theta \times H_\varphi| = -\frac{\omega k^3}{16\pi^2\varepsilon_0 r^2}\sqrt{\frac{\mu_0}{\varepsilon_0}}\sin^2\theta |p^2|$$

for the local time derivative of this momentum we obtain

$$\rightarrow \frac{d}{dt}p_r = -\frac{2c}{r}p_r$$

This time derivative of the linear momentum can be considered as an inertial force which is related to the kinetic energy of the propagating photon. This inertial force has a radial direction and therefore corresponds to a centripetal force giving rise to an angular momentum of the photon. For the angular momentum we obtain the relation:

$$\omega L = \iiint_{\theta=0,\varphi=0,r'=0}^{\frac{\pi}{2},2\pi,\infty} r\sin^2\theta \frac{d}{dt}p_r r^2 \delta(r'-r) d\varphi\, d\theta\, dr' = \frac{-\omega k^3}{8\pi^2 c\varepsilon_0}|p^2|\int_{\varphi=0}^{2\pi}\int_{\theta=0}^{\frac{\pi}{2}}\sin^4\theta\, d\varphi\, d\theta$$

$$\omega L = -\frac{\omega k^3}{32\, c\varepsilon_0}|p^2|$$

$$(14) \quad L = \frac{k^3}{32\, c\varepsilon_0}|p^2|$$

Considering that the angular momentum of the photon is quantized we obtain, by setting $L = \frac{h}{2\pi}$, the result:

$$(15) \quad |p^2| = \frac{32\, c\varepsilon_0 h}{2\pi k^3}$$

Replacing $|p|$ in the in the expression for the electric surface charge we obtain the relation:

$$(16) \quad q_e = 4\sqrt{\frac{2c\varepsilon_0 h}{\lambda}} = 4\sqrt{2\varepsilon_0 h\nu}$$

We can now compare this electrical surface charge of the electrical monopole to an electrical surface charge $q_{e;loop} = \frac{q_m}{c}$ of an equivalent electronic loop current of the magnetic monopole :

$$(17) \quad q(e; loop) = \sqrt{\frac{2c\varepsilon_0 h}{\lambda}} = \sqrt{2\varepsilon_0 h\nu}$$

So we see that the electrical surface charge of the circling loop current of the magnetic monopole of the magnetic photon is different from the electric surface charge of the electric monopole of the electric photon of the same wavelength.

Furthermore, we can say:

$$(18) \quad \frac{q_e(\lambda_1)}{q_e(\lambda_2)} = \frac{\sqrt{\nu_1}}{\sqrt{\nu_2}} = \frac{\sqrt{\lambda_2}}{\sqrt{\lambda_1}}$$

The electromagnetic properties of the magnetic photon and of the electric photon are summarized in table 1:

|  | Photon energy | Photon mass | Electric surface charge | Magnetic surface charge | Spin angular momentum |
|---|---|---|---|---|---|
| Magnetic photon | $h\nu$ | $\frac{h\nu}{c^2}$ | $\sqrt{\frac{2c\varepsilon_0 h}{\lambda}} = \sqrt{2\varepsilon h\nu}$ | $\sqrt{\frac{2\,h\nu}{\mu_0}}$ | $\frac{h}{2\pi}$ |
| Electric photon | $h\nu$ | $\frac{h\nu}{c^2}$ | $4\sqrt{2\varepsilon_0 h\nu}$ |  | $\frac{h}{2\pi}$ |

**Table 1** *electromagnetic properties of photons of frequency ν.*

**A derivation of the electromagnetic origin of the quantized charge of the electron**

With these results we can obtain an understanding of the inelastic electron positron pair production scattering process of an X- ray photon of a wavelength smaller than half the Compton wavelength $\lambda_c$

$$\lambda_{x-ray} \leq \frac{\lambda_c}{2}$$

and the generation of a charged particle from the electromagnetic energy of a photon. In the limiting case, $\lambda_{x-ray} = \frac{\lambda_c}{2}$, the magnetic photon has an equivalent electric surface charge

$$q_e = 2\sqrt{\frac{c\varepsilon_0 h}{\lambda_c}}$$

$$\frac{q_e^2}{c\varepsilon_0 h} = \frac{4}{\lambda_c}$$

One may ask the question of how the surface charge of the photon of an energy $2m_{0;electron}\ c^2$ can be related to the quantized charge of the electron or a positron of an energy $m_0\ c^2$. We look for an answer to this question by assuming that the static electromagnetic energy of the photon is converted to electrostatic energy of the electron and the positron. In order that the electrostatic energy of the electron is not infinite we assume that the electron is a circular homogeneously charged particle which has an electrostatic potential energy $U_{electron}$

$$U_{electron} = \frac{e^2}{\varepsilon_0 d} = eV_{electron} = m_0 c^2 = h\nu$$

This energy is set equal to half the dipolar electrical energy content of the photon of energy $2m_0c^2$

$$U_{photon} = \frac{1}{2\varepsilon_0} q_e^2 = 2h\nu = 2eV_{electron}$$

with $V_{electron} = 0{,}511$ MV

This leads to the relations

$$\frac{e^2}{\varepsilon_0\ d} = \frac{1}{4\varepsilon_0} q_e^2 = eV_{electron}$$

$$(19)\ e = \frac{ch}{\lambda_c V_{electron}}$$

$$(21)\ d = \frac{e^2}{\varepsilon_0 h\nu} = \alpha\lambda_c$$

where $\alpha = \frac{e^2}{\varepsilon_0\ hc}$ is the fine structure constant.

We come thus to the conclusion that we can derive the quantized charge of the electron from the electromagnetic field distribution of a photon. Furthermore we deduce from our electrodynamic model of a photon a radius of the electron, which is much smaller than the wavelength of an electron at rest or the wavelength of the photon from which the electron positron pair is generated which can be considered as a measure of the size of the photon. One may conclude that in the process of the generation of an electron positron pair the electromagnetic energy of the X-ray photon condenses to charged particles of a radius $\alpha\lambda_c$. A clear physical meaning can be assigned to the fine structure constant $\alpha$ as the ratio between the radius of the electron and its wavelength:

$$\alpha = \frac{d}{\lambda_c}$$

**Outline of a quantum topology of potential (cause)- and kinetic (effect)- energy of charge, mass and radiation as a unified quantum theory of electromagnetism and gravitation.**

P.A.M Dirac suggested that the existence of a magnetic monopole leads to a possibility to derive the quantized charge of the electron and ultimately to an alternative theory of quantum electrodynamics [6]. After having derived the quantized charge and mass of the electron in a way, which was inspired by the considerations of Dirac about the magnetic monopole, we follow his further suggestion to derive an alternative of quantum electrodynamic description of a photon, but we want to extend this concept to develop a unified theory gravitation and electrodynamics as not only a surface charge but also a mass can be attributed to the photon. It seems, that in the framework of Maxwell's theory a unification of gravitation and electromagnetism is not evident as was pointed out by Maxwell himself [7]. There is a long tradition of the search of higher dimensional unified theories of electromagnetism and gravitation which is documented extensively by D. Wünsch [8]. A unified quantum theory of electromagnetism and gravitation was specifically suggested on the basis of the concept of the photon as a solitary wave with a magnetic monopole by Sze Kui Ng [9], which did not request a higher dimensional space than the 3-D space. Our approach is a bit different as we make a distinction between magnetic and electric photons and we look for an alternative derivation of such a unified quantum theory of electromagnetism and gravitation on the basis of classical electromagnetic properties of a photon. We resort to a suggestion for a unified theory of electromagnetism and gravitation according to Bernhard Riemann. He submitted an article on a theory of electromagnetism from the Poggendorf Annalen der Physik und Chemie, which he, however, withdrew from publication. But the paper was published after his death in Riemanns collected scientific works edited by H. Weber and Dedekind [10]. Maxwell, who got to know of this paper, remarked, that the theory cannot be correct [11]. But Riemann spent much additional effort on the development of a unified theory of electromagnetism and gravitation. There is no publication on his unified theory. But in an addendum "Fragments of philosophical content" [12] to the mentioned collected scientific works of Riemann one finds a rather detailed mathematical elaboration of the unified theory with three sets of non linear partial differential equations. One also finds a few very clear sentences about the basic principles of his theory as was pointed out by D. Wünsch [8]. A translation of these sentences into English language reads as follows.

> "The effects of ponderable matter on ponderable matter are:
> 1) attractive and repulsive forces inversely proportional to the square of the distance
> 2) light and radiating heat.
> Both classes of phenomena may be explained if one assumes, that the whole infinite space is filled with a uniform substance and each particle of this substance acts directly on its environment.
> The mathematical law by which this occurs can be thought to be devided into
> 1) the resistance, by which a particle withstands a change in volume
> 2) the resistance, by which a line element withstands a change in length.
> On the first part gravitation and electrostatic attraction and repulsion is based, on the second the propagation of light and heat and the electrodynamic and magnetic attraction and repulsion."

We think that our explanation of the electron positron pair generation from a photon can be regarded as a condensation of Riemann's physical space consisting of a "uniform substance" to the elementary particles electron and positron. The second of the mentioned mathematical laws seems to be consistent with the special theory of relativity, as moving from one inertial reference frame to another leads to a distortion of a length unit. It seems to be a worthwhile effort to derive a unified quantum theory of electromagnetism and gravitation on the basis of the considerations and the system of non linear differential equations of B. Riemann.

**Acknowledgements**

This work was performed within a project "Edge plasmon mediated tip enhanced spectroscopy" Fi 608 of the German Science Council DFG.

I acknowledge the support by a short term mission and travel support by the COST Action MP1302 Nanospectroscopy of the European Union.

I acknowledge numerous discussions and collaborations with T. Grosjean, Y. Lefier , K. Tanaka and T.Maletzky which gave rise to the concepts which were elaborated in this contribution. An attempt to interpret results of experimental investigations performed in a cooperation with F. Fontein and H. Fuchs led to the concept of the photon as a solitary electromagnetic wave with a spin.